%% file: main.tex
\documentclass[sigconf]{acmart}

\usepackage{float}
\usepackage{subfigure}
\AtBeginDocument{%
  \providecommand\BibTeX{{%
    \normalfont B\kern-0.5em{\scshape i\kern-0.25em b}\kern-0.8em\TeX}}}
\usepackage{tabularx}
\usepackage{bbding}
\usepackage{graphicx}
\usepackage{geometry}
\usepackage{subfigure}
\usepackage{amsmath}
\usepackage{makecell}
\usepackage{float}
\usepackage{color}
\usepackage{booktabs}
\usepackage{caption}
\usepackage{tabu}
\usepackage{hyperref}
\usepackage{xcolor} 
\usepackage[normalem]{ulem} 

\copyrightyear{2026}
\acmYear{2026}
\setcopyright{cc}
\setcctype{by}
\acmConference[CHI '26]{Proceedings of the 2026 CHI Conference on Human Factors in ComputingSystems}{April 13--17, 2026}{Barcelona, Spain}
\acmBooktitle{Proceedings of the 2026 CHI Conference on Human Factors in Computing Systems (CHI'26), April 13--17, 2026, Barcelona, Spain}
\acmPrice{}
\acmDOI{10.1145/3772318.
3791863}
\acmISBN{979-8-4007-2278-3/
2026/04}

\usepackage{ulem}
\usepackage{xcolor}  

\begin{document}

\author{Suifang Zhou}
\email{sfzhou3-c@my.cityu.edu.hk}
\orcid{0000-0001-5103-580X}
\affiliation{%
\institution{City University of Hong Kong}
\city{Hong Kong}
\country{China}}

\author{Qi Gong}
\email{gongqi_77.0818@sjtu.edu.cn}
\orcid{0000-0002-5621-2673}
\affiliation{%
\institution{Shanghai Jiao Tong University}
\city{Shanghai}
\country{China}}

\author{Ximing Shen}
\email{ximing.shen@kmd.keio.ac.jp}
\orcid{0000-0003-4272-0068}
\affiliation{
\institution{Keio University \\Graduate School of Media Design}
\city{Yokohama}
\country{Japan}}

\author{RAY LC}
\authornote{Correspondences can be addressed to ray.lc@cityu.edu.hk.}
\email{ray.lc@cityu.edu.hk}
\orcid{0000-0001-7310-8790}
\affiliation{
\institution{City University of Hong Kong\\Studio for Narrative Spaces}
\city{Hong Kong, SAR}
\country{China}}

\renewcommand{\shortauthors}{Zhou, et al.}

\title[Tell Me What I Missed]{Tell Me What I Missed: Interacting with GPT during Recalling of One-Time Witnessed Events}


\begin{abstract}

LLM-assisted technologies are increasingly used to support cognitive processing and information interpretation, yet their role in aiding memory recall—and how people choose to engage with them—remains underexplored. We studied participants who watched a short robbery video (approximating a one-time eyewitness scenario) and composed recall statements using either a default GPT or a guided GPT prompted with a standardized eyewitness protocol. Results show that default-condition participants who believed they had a clearer understanding of the event were more likely to trust GPT’s output, whereas guided-condition participants showed stronger alignment between subjective clarity and actual recall. Additionally, participants evaluated the legitimacy of the individuals in the incident differently across conditions. Interaction analysis further revealed that default-GPT users spontaneously developed diverse strategies, including building on existing recollections, requesting potentially missing details, and treating GPT as a recall coach. This work shows how GPT–user interplay subconsciously affects beliefs and perceptions of remembered events.

\end{abstract}

\ccsdesc[500]{Human-centered computing~Empirical studies in collaborative and social computing}

\keywords{Large Language Models, Artificial Intelligence, memory}

\begin{teaserfigure}
    \centering
    \includegraphics[width=1\linewidth]{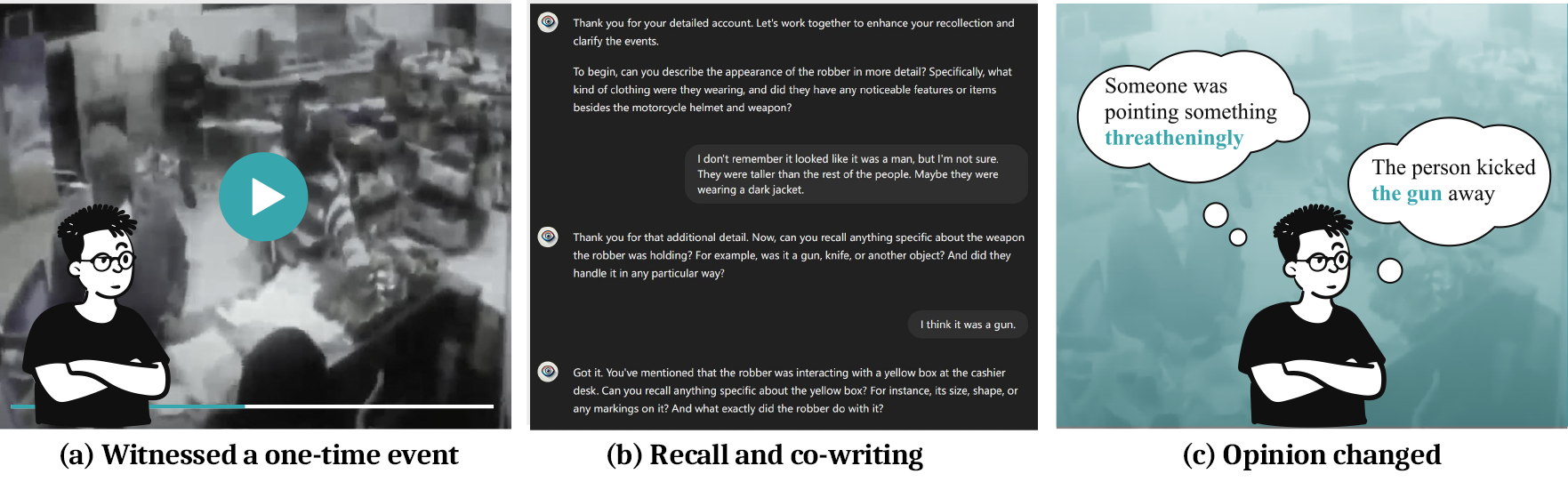}
    \caption{An illustration of the study workflow. (a) Participants watched a video about a suspected robbery incident. (b) Participants wrote a statement about what happened in the video with a GPT writing assistant. (c) We measured participants' self-perceived incident recognition, their opinions about the incidents and GPT, and inquired about how they interact with the GPT.}
    \Description{A series of figures illustrating the study workflow is presented, from left to right they are labelled as A, B, and C. (a) Participants watched a video about a suspected robbery incident. (b) Participants wrote a statement about what happened in the video with a GPT writing assistant. (c) We measured participants' self-perceived incident recognition, their opinions about the incidents and GPT, and inquired about how they interact with the GPT.}
    \label{fig:teaser}
\end{teaserfigure}
\maketitle

\section{Introduction}\label{sec:Introduction}
\input{sections/01-Intro.tex}

\section{Background}\label{sec:Background}
\input{sections/02-Background.tex}

\section{Method}\label{sec:Method}
\input{sections/03-Method.tex}

\section{Result}\label{sec:Result}
\input{sections/04-Result}

\section{Discussion}\label{sec:Discussion}
\input{sections/05-Discussion.tex}

\section{Conclusion}\label{sec:Conclusion}
\input{sections/06-Conclusion.tex}

\section{Appendix}\label{sec:Appendix}
\input{sections/Appendix.tex}


\bibliographystyle{ACM-Reference-Format}
\bibliography{references}

\end{document}

%% file: sections/01-Intro.tex
Understanding the interplay between human cognition and interaction processes in human–LLM collaboration—how people can use these systems reliably and effectively, and what vulnerabilities accompany their use—has become an urgent practical challenge \cite{NIST_AI_RMF_2023, xu2023transitioning}. This urgency is driven by the rapid integration of LLM-powered tools into everyday life and, increasingly, into established professional workflows where precise factual documentation is critical, including medicine, journalism, legal drafting, and policing contexts such as investigative report write-ups and eyewitness statements \cite{Wang_2022, Kitamura_2023, Broussard_2019, Phelps_2022}. While these systems can enhance efficiency and consistency, recent work cautions that they may also introduce subtle distortions or biases, raising concerns about the authenticity and fairness of the information they help produce \cite{Castillo-Campos_2024}.

In line with this need, researchers have examined how LLM design features shape user behavior and outcomes in human–AI collaboration, particularly when such effects raise safety and reliability concerns. A recurring theme is how LLMs can influence users’ perceptions, intentions, and decisions \cite{LeeSee_2004,LoggMinsonMoore_2019,DietvorstSimmonsMassey_2015}. For instance, role framing and agent personas—often instantiated through prompting—can shift users’ trust, emotional responses, and how they engage with and interpret AI-generated content \cite{Pataranutaporn_2023}. AI-generated explanations, especially when presented with authoritative cues, can increase persuasiveness and facilitate misinformation uptake, shaping users’ beliefs \cite{danry2025deceptive}. In complex tasks such as dilemma decision-making and sensemaking, users with higher uncertainty or lower confidence in their own judgments may defer more readily to LLM advice than to human guidance \cite{Gunaratne_2018, Logg_2019}, whereas users who perceive themselves as highly competent may discount AI suggestions—even when reliance would improve performance \cite{He_2023}.

Interest has also grown in how AI-mediated tools influence memory processes \cite{huang2024unavoidable, sun2024ai, chan2024conversational}. Memory is a foundational cognitive function and is especially consequential in tasks requiring accurate factual documentation, given its tight coupling with language and meaning through schemas, contextual cues, and semantic associations \cite{TulvingThomson_1973, BrewerTreyens_1981,Freire_2024}. Recent advances in large language models (LLMs) heighten these concerns because LLMs intervene directly in language—the primary medium through which memories are articulated and formalized \cite{kim2023chatgpt}. By supplying phrasing, categories, and narrative structure during writing, and by occasionally introducing plausible but inaccurate content, LLM tools may shape which retrieval cues are activated and how ambiguous details are resolved during reconstruction \cite{Ji_2023_SurveyHallucination,banerjee2025llms}. This is particularly salient when LLMs assist legal documentation, where even minor distortions can misdirect investigations, undermine testimony, or affect judicial outcomes \cite{Taylor_2011, DeRenzo_2020}.

Eyewitness statement writing is therefore a particularly consequential setting for examining these risks. Witnesses must translate brief and often stressful observations into written narratives under uncertainty, drawing on incomplete memory traces, inferences, and post-event information \cite{buckhout1974eyewitness,reno2001eyewitness,loftus2019eyewitness}. Moreover, because reconstructive recall is strongly cue-dependent, producing a coherent narrative places heavy demands on linguistic cues and semantic associations to organize retrieval and fill gaps \cite{TulvingThomson_1973,Bartlett_1932}. As many witnesses are laypeople with limited familiarity with legal writing and reporting conventions, they may be especially inclined to rely on LLM-powered tools for assistance \cite{Fiederer_2019_TrustConfidence, weissenbergerrule}. Yet when LLMs contribute to statement writing by suggesting wording, structure, or details, they may not only aid articulation but also shape attention, steer ambiguity resolution, and influence what witnesses ultimately accept as having occurred \cite{chan2024conversational, OpenAI_WhyHallucinate_2025, banerjee2025llms}. Consequently, designing and configuring such tools to support accuracy and reliability is a critical yet underexplored challenge.

Against this backdrop, while a growing body of work cautions that exposure to AI-generated content and engagement with AI-mediated tools can introduce memory distortions or yield information that fails to meet authenticity requirements \cite{Taylor_2011, DeRenzo_2020}, comparatively little research has examined the interactional processes and mechanisms through which such failures emerge. In particular, we still lack a grounded understanding of how people practically engage with LLMs during memory-related tasks, and how design feature—such as prompting structures—when enacted in interaction, shape the memory processes involved in retrieving, interpreting, and reconstructing information. Moreover, given that human memory is inherently malleable and susceptible to subjective influence—and that prior findings show AI-mediated tools can shape users’ perceptions, interpretations, and subsequent judgments—a critical next step is to investigate how collaboration with an LLM may steer the ways individuals reconstruct and make sense of what they believe they remember when such systems are embedded in memory-related activities \cite{loftus2019eyewitness,Loftus_2005,Pataranutaporn_2024_SyntheticHumanMemories}.

In response to this gap, we designed a study in which participants interacted with a GPT-based “witness assistant,” guided by standardized eyewitness investigation protocols \cite{reno2001eyewitness}. This setup was intended to emulate a speculative yet contextually grounded use case, in which individuals employ LLM-powered tools to assist in composing testimonial narratives within evidence-based investigative scenarios. Within this context, we pose the following research questions:


\textbf{RQ1:} \textit{How do different LLM prompting designs affect people’s ability to recall witnessed events?}

\textbf{RQ2:} \textit{How do different prompting designs shape post-hoc perceptions of a witnessed event?}

\textbf{RQ3:} \textit{How do users utilized the LLM-powered tools for writing witnessed statement?}  


Informed by established findings that GPT-based systems can amplify false memories in eyewitness interviews \cite{pataranutaporn2024synthetic,pataranutaporn2025slip}, we extend this line of inquiry by examining how different forms of prompting with the same LLM shape the mechanisms of that influence, specifically by observing how the model affects users’ internalization processes under comparable LLM agent. To address this, we introduce two conditions: a guided interaction group, in which the GPT agent was prompted using a standardized eyewitness preliminary protocol \cite{reno2001eyewitness}, and a natural interaction group, in which participants used the default GPT without structured prompting. This design enables us to investigate how varying forms of LLM involvement—structure-tuned cueing versus free generative suggestion—shape users’ perceptions of a witnessed event and, ultimately, what they believe they remember.

We employ this all-LLM experimental setup without including a conventional control group, such as a human-led witness recall condition, to isolate potential interactional effects under controlled conditions. Because the study centers on differences in GPT-based interaction procedures rather than human–AI comparisons, this approach minimizes external sources of variation—such as interviewer behavior, procedural differences between human- and AI-led interviews, and interviewer-specific characteristics—thereby ensuring greater consistency and comparability across conditions. Holding these factors constant enables more direct isolation of the effects of LLM involvement on memory-related performance.

We recruited 28 participants through our social network. Each participant viewed a 36-second surveillance video depicting an indoor invasion and robbery. After viewing the video, participants waited 15 minutes to allow for initial cognitive processing of the event. They were then instructed to use GPT to assist them in composing a detailed written statement of what they remembered from the scene. Participants were randomly assigned to either the guided or natural interaction condition.

After completing the statement-writing task, participants waited one hour before filling out a post-experiment questionnaire consisting of three sections: (1) factual item identification based on the video, (2) subjective interpretation—including perceived intensity, emotional impact, clarity, and judgments of legitimacy toward the event, and (3) evaluation of GPT’s performance during the recall-writing process, including perceived helpfulness and confidence in the tool. Finally, we conducted semi-structured interviews to further investigate participants’ interaction behaviors and their perceptions of GPT’s role and usefulness during the recall task.

Our results show that self-perceived memory clarity was associated with perceptions of GPT that affected the recall of factual events. We also found that participants in the default GPT group exhibited significantly greater disfavor toward the intruder in the incident. This further suggests that participants’ perceived legitimacy of the factual event was amplified in the natural interaction condition. We make the following contribution: (1) we demonstrate the fragility of human memory and perception under GPT's influences, (2) we gain empirical insights about how people interact with purposeful GPT for recording a factual event, where they attempt to use GPT to improve writing efficiency while ensuring the authenticity and accuracy of the content, (3) we discuss implications for the use of GPT in evidence-based professional fields such as medical writing, legal investigation and journalism.


%% file: sections/02-Background.tex
\subsection{LLMs for Factual Documentation in Investigative Context}

LLMs are increasingly adopted to support documentation and other writing-intensive tasks \cite{Cambon_2023, yang_ai_2022, Silva_2024, chen_once_2025}. Prior HCI work shows that conversational AI can help people produce and refine text by offering structural scaffolding, expanding ideas, and improving coherence across everyday writing contexts such as journaling and summarization \cite{Kim_2024, fu_being_2024, zeng_ronaldos_2025, He_Houde_2024, Goyal_2023}. These capabilities have motivated exploration of LLM-powered tools in domains where written accounts of factual events are central, including journalism, healthcare, and legal practice \cite{Wang_2022, Freire_2024, Phelps_2022}.

This momentum has begun to extend into law-enforcement documentation, where written narratives may later function as evidence \cite{minhas2022protecting,dando2025collecting}. Early deployments suggest that conversational agents can streamline portions of investigative workflows and lower barriers to public reporting, while commissioned reports argue they may also reduce administrative burden and help users notice or correct issues that static forms can miss \cite{hepenstal2021developing, van2022source}. Recent research further indicates that instruction-tuned LLMs can support the scalable analysis of police incident narratives: Relins et al. \cite{relins2025using} evaluated instruction-tuned LLMs for deductive coding of Boston Police Department incident narratives and found that agreement with human coding improved with larger models and tailored prompting, pointing to hybrid pipelines in which LLMs screen cases and humans review ambiguous instances.

At the same time, in these evidence-based domains, efficiency cannot come at the expense of rigor, authenticity, or fairness \cite{Kovach_2014, Foreman_2009, Oates_2002, Taylor_2011, DeRenzo_2020}. Emerging evidence suggests particular risks in eyewitness contexts, because witnesses are typically non-experts and—amid uncertainty, stress, and high perceived AI usefulness—may be especially prone to over-rely on AI assistance when articulating what they saw \cite{bender2021dangers,weissenbergerrule,tomas2025chatbots}. For example, one study \cite{chan2024conversational} simulated crime witness interviews and compared multiple conditions using control, survey-based questioning, and chatbot-based interviews, finding that an LLM-powered generative chatbot increased false-memory reports relative to non-generative and control conditions, with some incorrect details persisting over time. These findings highlight a central challenge: although LLMs may streamline investigative documentation, their involvement in how witnesses recount events—through questioning, paraphrasing, or suggestion—may also shape what witnesses come to believe they remember, underscoring the need for safer interaction designs that support accurate documentation \cite{pataranutaporn2025slip}.

\subsection{Human Perception of Large Language Models}

A growing body of HCI research shows that LLM-powered tools can shape users’ perceptions and attitudes, not just the text they produce \cite{he_i_2025, zhou_eternagram_2024}. For example, Jakesch et al. \cite{Jakesch_2023} found that co-writing with an LLM writing assistant influenced both participants’ written statements and their reported attitudes, and Sharma et al. \cite{Sharma_2024} showed that an LLM-powered conversational search system can increase selective exposure and opinion polarization. As these tools are deployed in influence-sensitive contexts such as education, games, and social media \cite{Hedderich_2024, Kazemitabaar_2024, Han_2024, Qin_2024, Meier_2024}, understanding when and how they shape psychological outcomes has become a key design concern.

Crucially, these effects are not solely a property of model output; they are mediated by how the system is presented and interpreted in interaction. Users’ prior beliefs, expectations, and self-assessments shape how they evaluate and incorporate LLM responses, which can in turn reinforce (or undermine) those beliefs and expectations in a feedback loop \cite{Eslami_2018_FAT, LoggMinsonMoore_2019, TrustRelianceDevelopment_AIAvice}. Other work demonstrates that describing the same chatbot with different motives shifts perceived trustworthiness \cite{Pataranutaporn_2023}, and research on algorithm appreciation suggests that people may defer to algorithmic advice over human guidance under uncertainty \cite{Gunaratne_2018, Logg_2019}. Complementary qualitative work further underscores the importance of aligning LLM behavior with domain constraints and user risk perceptions, including when LLM advice is appropriate and how users navigate privacy and disclosure tradeoffs \cite{Cheong_2024, Zhang_2024}.

At the same time, as LLMs are increasingly treated as authoritative sources \cite{Kapania_2022, Araujo_2020}—and can outperform humans on some tasks \cite{Kim_Leonte_2024}—questions extend to agency and interpretation during collaboration. Draxler et al. \cite{Draxler_2024}, for instance, found that co-writing with LLMs can reduce users’ sense of ownership over the resulting content. Some users exhibit a greater tendency to follow advice from LLMs than from human advisors \cite{Gunaratne_2018, Logg_2019}, while others overestimate their own abilities and become reluctant to rely on AI tools, potentially impeding effective human-AI collaboration \cite{He_2023}.

\subsection{Human Memory and AI-Mediated Influences}

Decades of work in cognitive psychology conceptualize memory as reconstructive: what people encode depends on attention and appraisal at the time of perception, and later recall involves reassembling events from partial traces rather than replaying a veridical record \cite{Bartlett_1932, Schacter_1999}. This reconstruction is tightly coupled with language and meaning—schemas, contextual cues, and semantic associations help organize recall and fill in gaps \cite{TulvingThomson_1973, BrewerTreyens_1981}. Accordingly, memory is subjective and malleable: post-event information and suggestive wording can systematically shape what people report, sometimes increasing confidence in details that never occurred \cite{loftus1989misinformation}.

HCI has long examined how technologies mediate remembering, including memory-support tools for older adults \cite{Li_2019} and people with memory impairments (e.g., amnesia, dementia) \cite{Hodges_06, Wu_2010, Franklin_2021}. Prior systems include wearable reminders that surface identity cues in situ \cite{Jager_2011} and diary-based tools designed to scaffold episodic recall \cite{Le_2016, Li_2024}. At the same time, HCI research documents memory risks introduced by mediated experiences, such as source confusion between VR and reality \cite{Clinch_2021, Bonnail_2024, Bonnail_2023}. More recently, attention has turned to AI-mediated memory effects, particularly around AI-generated or AI-modified content that can blur boundaries between authentic and fabricated experiences \cite{corsi2024spread,Pataranutaporn_2024_SyntheticHumanMemories}.

LLM-powered tools sharpen these concerns because they operate directly on linguistic and semantic structure—the same cues that support memory reconstruction. Classic evidence shows that linguistic framing can alter event appraisal and later recall (e.g., the “smashed” vs. “hit” manipulation in eyewitness questioning) \cite{Loftus_Palmer_1974}. Given growing evidence that LLM interactions can propose plausible details and guide reflection through conversational prompts, a key concern is that the language these systems introduce may shape users’ beliefs about what occurred, which can then carry over into how the event is reconstructed in memory \cite{chan2024conversational, Loftus_2005, OpenAI_WhyHallucinate_2025, Bender_2021_StochasticParrots}. In this sense, LLM-assisted interactions may not only change how accounts are written, but also shift what users come to believe they remember. However, the mechanisms through which interactive LLM conversations shape memory formation and distortion remain comparatively underexplored. As LLM-based dialogue systems become embedded in routine information seeking, reflection, and documentation, clarifying their role in shaping—and potentially distorting—memory is increasingly urgent.

%% file: sections/03-Method.tex
\subsection{Participants}
A total of 28 participants took part in this study. All participants completed the full procedure, which included a writing task, a post-task survey, and a follow-up interview. Participants were recruited from a university-affiliated social network. Participants' English language proficiency levels were self-reported as follows: 12 identified as intermediate, 11 as advanced, and 5 as basic users. Familiarity with GPT-based tools also varied: 15 participants described themselves as very familiar, 5 as moderately familiar, 4 as extremely familiar, 2 as somewhat familiar, 1 as expert level, and 1 as slightly familiar. Regarding the frequency of GPT usage, 14 participants reported using GPT daily, 5 indicated very frequent use (a few times a week), and 4 reported using it frequently (once a week). The remaining participants used GPT sometimes (n = 1), occasionally (n = 1), or rarely (n = 3). Participants received a payment of 50 Hong Kong dollar upon completion of the experiment and surveys.

\begin{table*}[h!]
\centering
\resizebox{0.975\textwidth}{!}{%
\begin{tabular}{|p{1.5cm}|p{4.5cm}|p{4.5cm}|p{4.5cm}|}
\hline
\textbf{ID} & \textbf{English Level} & \textbf{Familiarity with GPT} & \textbf{Frequency of Use} \\ \hline
P1 & Advance     & Very familiar       & Very frequently \\ \hline
P2 & Advance     & Extremely familiar  & Frequently \\ \hline
P3 & Intermediate& Expert level        & Daily \\ \hline
P4 & Intermediate& Very familiar       & Daily \\ \hline
P5 & Basic       & Very familiar       & Daily \\ \hline
P6 & Advance     & Extremely familiar  & Very frequently \\ \hline
P7 & Basic       & Very familiar       & Daily \\ \hline
P8 & Basic       & Extremely familiar  & Very frequently \\ \hline
P9 & Advance     & Very familiar       & Sometimes \\ \hline
P10 & Intermediate& Moderately familiar & Frequently \\ \hline
P11 & Advance     & Moderately familiar & Occasionally \\ \hline
P12 & Intermediate& Very familiar       & Daily \\ \hline
P13 & Intermediate& Very familiar       & Daily \\ \hline
P14 & Advance     & Extremely familiar  & Daily \\ \hline
P15 & Intermediate& Very familiar       & Daily \\ \hline
P16 & Intermediate& Moderately familiar & Very frequently \\ \hline
P17 & Intermediate& Somewhat familiar   & Daily \\ \hline
P18 & Intermediate& Very familiar       & Daily \\ \hline
P19 & Basic       & Very familiar       & Frequently \\ \hline
P20 & Advance     & Moderately familiar & Rarely \\ \hline
P21 & Basic       & Very familiar       & Very frequently \\ \hline
P22 & Advance     & Very familiar       & Frequently \\ \hline
P23 & Advance     & Very familiar       & Daily \\ \hline
P24 & Advance     & Somewhat familiar   & Rarely \\ \hline
P25 & Intermediate& Moderately familiar & Daily \\ \hline
P26 & Advance     & Slightly familiar   & Rarely \\ \hline
P27 & Intermediate& Very familiar       & Daily \\ \hline
P28 & Intermediate& Very familiar       & Daily \\ \hline
\end{tabular}
}
\caption{Participant Demographics (N=28)}
\Description{Summary of 5 participants in the formative interview phase.}
\end{table*}

\begin{itemize}
    \item Consent: We ensured that participants provided informed consent for the study. The consent form comprehensively outlined the research's objectives, procedures, potential risks, and benefits. It emphasized the voluntary nature of participation and the participants' right to withdraw at any time.
    \item Anonymity: Participant privacy and anonymity were rigorously protected, and no personally identifiable information was collected. We took precautions to avoid gathering any sensitive data that could be used to identify individuals.
    \item IRB: University ethics review board approved the human subject testing in this project.
    \item Ethic: We adhered to fundamental ethical principles in our study, encompassing the respect of participants' rights, minimization of risks, and the assurance of data confidentiality and security.
\end{itemize}

\subsection{Study Design and Procedure}

Our primary goal is to understand how participants employ LLM-powered tools to compose testimonial narratives in evidence-based investigative scenarios, and how collaboration with an LLM may steer the ways individuals reconstruct and make sense of what they believe they remember. To examine these processes, we implemented two distinct GPT-based interaction conditions. The first was a guided interaction condition, in which the LLM was prompted using a standardized eyewitness investigation protocol \cite{reno2001eyewitness}. This setup reflects a speculative yet contextually grounded scenario aligned with prior work that has emulated structured uses of LLMs in investigative contexts. The second was a natural interaction condition, in which participants used the default GPT interface—capturing the unstructured, everyday way people routinely engage with LLMs without formal guidance. This condition represents the most common use case of GPT: free-form, unguided interaction characteristic of real-world documentation practices.

We adopted this two-condition design—rather than comparing LLM-assisted interviews with human-led investigations—because prior studies have already shown that LLMs can meaningfully alter, and at times distort, memory in investigative settings \cite{pataranutaporn2024synthetic,pataranutaporn2025slip}. Building on this foundation, we intentionally maintain an all-LLM experimental environment to support a streamlined and controlled procedure. This approach allows us to narrow external sources of variation, such as interviewer behavior or interaction norms, and keep the procedural elements consistent across conditions, while focusing specifically on how variations in LLM involvement itself influence memory-related performance under comparable operational conditions.

Participants were randomly and assigned to one of two GPT conditions via a designated link: the Natural Interaction condition (P1–P13, n = 13) or the Guided Interaction condition (P14–P28, n = 15). Participants were not informed of their condition assignment. All necessary instructions for the experiment would be provided through the link and displayed on-screen. To ensure procedural compliance and enable process observation, participants were encouraged to enable screen sharing and video recording during the whole study. Following this, all participants completed a standardized video-viewing task, which involved watching a 36-second surveillance recording depicting an indoor invasion and robbery. 

The 36-second surveillance video depicts an ambiguous incident in a convenience store involving multiple individuals. A helmeted person enters, and a rapid sequence of events unfolds: people collapse, a yellow liquid spills, and a bottle rolls across the floor. The lack of clearly visible weapons or explicit actions creates uncertainty about what actually occurred—mirroring the ambiguity typical of real eyewitness scenarios. The video contains no graphic content such as visible violence, blood, or other distressing elements. The complete video is available via the link provided in the footnote.\footnote{Video link: [https://www.youtube.com/watch?v=pqBt7P0LW4M]}. Participants were instructed to watch the video only once before proceeding to the writing task.

\setlength{\intextsep}{20pt} 
    \begin{figure*}[h]
        \centering
        \includegraphics[width=0.98\linewidth]{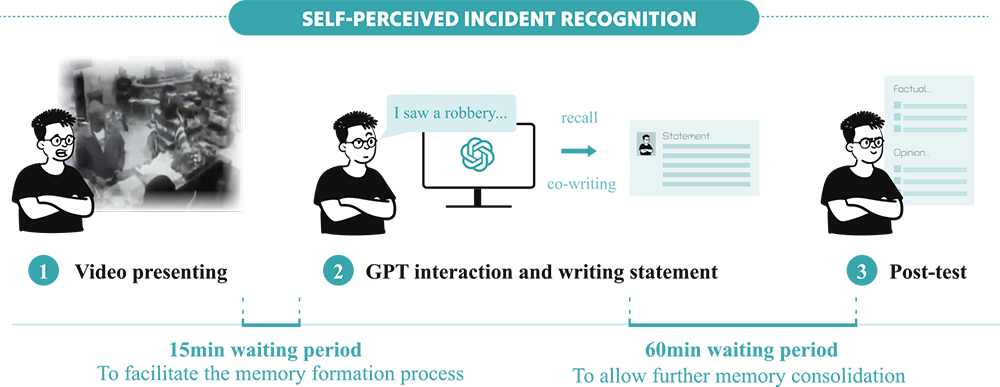}
        \caption{An illustration of the procedure of the study. (1) Participants watched the video, and waited for 15 minutes. (2) Participants used the GPT witness assistant to compose a statement describing what they remembered about the video, and then waited for 60 minutes. (3) Participants filled in the questionnaire that includes a factual items identification section, a subjective interpretation section and interview.}
        \label{fig:ExperimentalProcedure}
        \Description{An illustration of the procedure of the study. (1) Participants watched the video, and waited for 15 minutes. (2) Participants used the GPT witness assistant to compose a statement describing what they remembered about the video, and then waited for 60 minutes. (3) Participants filled in the questionnaire that includes a factual items identification section, a subjective interpretation section and interview.}
    \end{figure*}

The primary objective was to understand the effect of the interaction with prompted GPT on memory of a witnessed incident. To facilitate the memory formation process, a 15-minute waiting period was implemented immediately after the video-watching task. This interval serves two purposes: it simulates the realistic delay between witnessing an event and composing a formal statement, and it has been shown that brief wakeful rest periods (10–15 minutes) can enhance memory retention by reducing interference \cite{dewar2012brief}. During the waiting period, participants could engage in passive activities (e.g., resting, eating, drinking) but were explicitly prohibited from consuming any informational content—such as videos, books, news, or social media—that could interfere with memory consolidation. Participants were instructed to return to the experiment exactly after 15 minutes.

Following the waiting period, participants proceeded to the witness statement writing task. In both conditions, participants received identical instructions:``You are an eyewitness to the incident shown in the video. Please use GPT-4 to help you compose a detailed witness statement describing what you observed. You may interact with GPT-4 in whatever way you find most helpful to create an accurate and complete account.''

While participants in both conditions used the same GPT interactive interface, the chat agent was configured with different prompts tailored to each condition. Participants were blind to their condition assignment. To simulate natural usage scenarios, participants were allowed to interact with GPT-4 in various ways according to their usage habits. They could use GPT-4 to recall event details, improve clarity, or edit generated text as needed. No requirements were imposed on interaction duration or number of rounds. Participants had full autonomy to decide when their statement was complete and ready for submission.

After completing the statement writing task, participants underwent a second 60-minute waiting period to allow for further memory consolidation, as synaptic consolidation processes have been shown to stabilize within approximately one hour following encoding \cite{dudai2004neurobiology}. The instructions during this interval mirrored those of the initial waiting period, aiming to minimize external influences on the memory formation process. Following this delay, participants proceeded to the post-task survey, which included factual recall questions, subjective interpretation assessments, and demographic data collection. Given the linguistic demands of the task and the practical application of GPT, we also gathered information on participants’ English language proficiency, prior experience with GPT, and frequency of GPT usage. At the conclusion of the experiment, participants took part in a semi-structured interview designed to explore how they approached the GPT witness assistant when composing and documenting their statements about the robbery video. 

\subsection{Condition and Prompt Engineering}

\setlength{\intextsep}{20pt} 
    \begin{figure*}[h]
        \centering
        \includegraphics[width=0.93\linewidth]{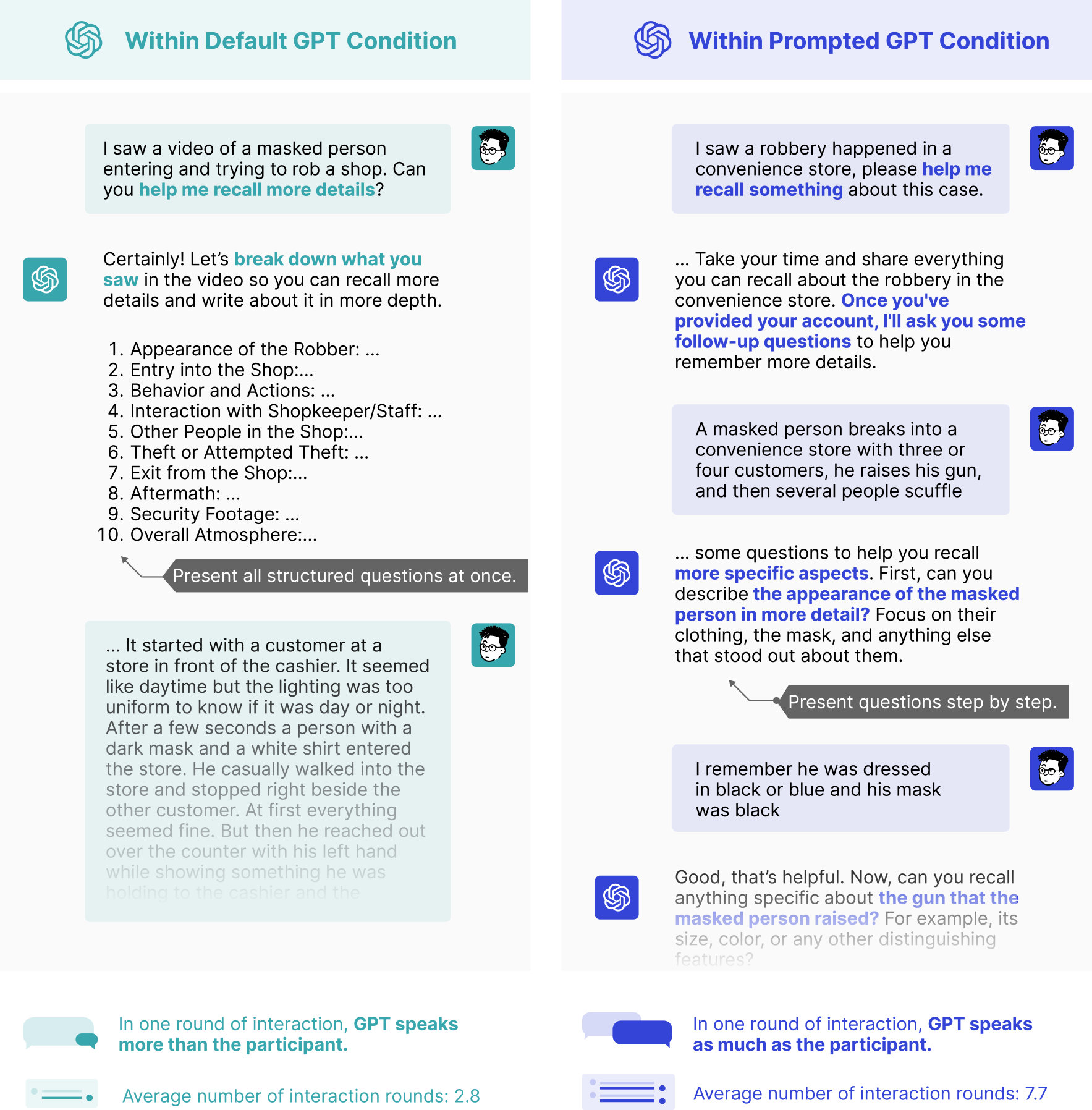}
        \caption{Examples of how participants from each group interact with GPT. Left: In the Default GPT Condition group, participants interacted with the GPT naturally. Right: In the Prompted GPT Condition group, participants interacted with the prompted GPT that was configured with specific prompts derived from a guidance on the preliminary investigation.}
        \label{fig:ExperimentalProcedure}
        \Description{Two examples of how how participants from each group interact with GPT are displayed side by side. Left: In the Default GPT Condition group, participants interacted with the GPT naturally. Right: In the Prompted GPT Condition group, participants interacted with the prompted GPT that was configured with specific prompts derived from a guidance on the preliminary investigation.}
    \end{figure*}

\subsubsection{Natural Interaction GPT Condition (default condition):} Participants in this condition engaged with a web-based user interface utilizing an unprompted GPT-4 model to prepare a witness statement. The objective was to observe the participants' interaction with the GPT-4 model and assess the impact of such interactions on memory recall in common, real-world scenarios without deliberate intervention. In this context, GPT-4 served as a linguistic processing assistant and has all the default feature that GPT-4 has, facilitating tasks such as statement composition by assisting with the organization of thoughts and language through natural, unstructured interaction with the default GPT-4 chatbot.

\subsubsection{Prompted Interaction GPT Condition (prompted condition):} In this condition, participants were tasked with the same instruction on preparation of a witness statement; however, the GPT-4 model was configured with specific prompts derived from a guidance on the preliminary investigation process \cite{reno2001eyewitness}. These prompts were crafted to elicit open-ended responses during the interaction, encouraging participants to elaborate on the details of the incident based on their initial input. For instance, if a participant mentioned a vehicle, the chat agent would inquire further with questions such as, "What color was the vehicle?" while carefully avoiding leading questions like, "Was the vehicle red?" Each follow-up question was generated from the participant’s previous responses, functioning analogously to a legal investigator meticulously reviewing and probing each element of the witness’s statement within the broader context of the statement formulation process. Both conditions used GPT-4.0 to ensure consistency in model capabilities.

\subsection{Measurement}

\subsubsection{Memory Recall Accuracy}
Following the main experiment, a post-experiment assessment was conducted to evaluate the accuracy of participants’ memory recall. The assessment was based on content from the surveillance video. A factual memory recognition test consisting of 50 questions was administered, with each question derived from specific visual details observed in the video. The questions followed a forced-choice format with three possible responses: "yes," "no," or "unsure." Participants’ answers were scored cumulatively, with a maximum possible score of 50. To facilitate analysis, the test was divided into five thematic categories, each comprising 10 items:

\begin{itemize} 
\item Incident Scene: Assessing the accuracy of recalling the environment where the incident occurred. 
\item Intruder's Profile: Assessing the accuracy of recalling the appearance and characteristics of the intruder. 
\item Intruder's Actions: Assessing the accuracy of recalling the intruder's behavior during the incident. 
\item Other Individuals' Profiles: Assessing the accuracy of recalling the appearance and characteristics of other individuals involved, such as customers and the cashier. 
\item Other Individuals' Actions: Assessing the accuracy of recalling the actions and reactions of other individuals during the incident. \end{itemize}

\subsubsection{Interpretation of the Witnessed Event}

To investigate the impact of GPT on participants’ subjective interpretation of the witnessed event, we administered a 14-item opinion-based survey. In contrast to the factual memory accuracy test, this section aimed to probe participants’ subjective dimensions—specifically, how they internalized and interpreted the incident after interacting with different GPT conditions. The items were designed to assess four key dimensions: perceived intensity (4 items), emotional responses (3 items), clarity of the incident (4 items), and judgments regarding the legitimacy of the individuals involved (3 items). All responses were recorded using a 7-point Likert scale, ranging from 1 (strongly disagree) to 7 (strongly agree).

\begin{itemize}
    \item Perceived Intensity: Refers to the level at which participants consider the incident to be violent; a higher score indicates a greater perceived intensity of violence.
    \item Emotional Response: Measures the emotional impact of the incident on participants; a higher score signifies stronger emotional arousal.
    \item Perceived Incident Clarity: Assesses how clearly and confidently participants believe they understand the incident they observed; a higher score reflects greater perceived clarity.
    \item Legitimacy Judgment: Examines participants' inclination toward those involved in the incident; a higher score indicates a more favorable view of the intruder.
\end{itemize}

\subsubsection{Self Perceived GPTs Accuracy}

We incorporated a measurement to assess the perceived impact of GPT in supporting memory recall and documenting task. This assessment included participants’ judgments regarding the system's reliability, their subjective confidence in its output, and its utility in aiding both memory retrieval and the articulation of a coherent witness statement. Each of these constructs was measured using a set of five items on a 7-point Likert scale, ranging from 1 to 7. The measurement focused on the following aspects:

\begin{itemize}
\item Perceived accuracy of GPT in supporting memory recall;
\item Confidence in GPT’s output during the interaction;
\item Perceived enhancement of memory retrieval facilitated by GPT;
\item Perceived improvement in the quality of the witness statement;
\item Overall usefulness of GPT as a tool in the context of evidence-based reporting.
\end{itemize}

\subsubsection{Witness Assistant Usage Pattern}

One objective of this study was to answer how participants engaged with the GPT for both conditions when composing statements about the robbery video. To this end, we conduct a semi-structured interview was conducted following the completion of the survey component. The interview aimed to gain insight into participants' interaction strategies, perceptions of authorship, and decision-making processes related to tool usage. Participants were asked the following questions in sequence:

\begin{itemize}
\item \textit{Q1: "How did you approach the witness assistant?"}
\item \textit{Q2: "How did you find the overall experience of using the provided method for this task? How did using the provided method affect your approach to the task compared to simply recalling?"}
\item \textit{Q3: "Do you feel you are the author of your writing?"}
\item \textit{Q4: "What influences your decision to conclude your interaction with the tool and submit your statement?"}
\end{itemize}

\subsection{Data Analysis}

We employed the reflexive thematic analysis method following \cite{Braun_2012} to analyze the data collected from the interview.
Initially, the primary researcher undertook the task of coding the video and audio transcripts using an inductive methodology. The preliminary codes were then deliberated upon and refined in collaboration with the second and third researchers. As the conceptualization of themes began to take shape, we expanded our discussions to include the fourth author, enabling further refinement of the themes.


%% file: sections/04-Result.tex
\subsection{Memory Recall Performance and Subjective Internalization of the Incident}

\setlength{\intextsep}{20pt} 
    \begin{figure*}[h]
        \centering
        \includegraphics[width=0.63\linewidth]{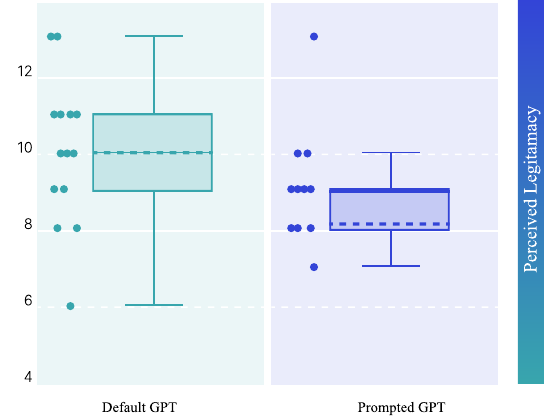}
        \caption{Box plot illustrating significant differences between Default GPT and Prompt GPT groups: participants in the Default GPT group showed greater disfavor toward the intruder and more favorable attitudes toward other individuals involved in the incident.}
        \Description{A box plot illustrating significant differences between Default GPT and Prompt GPT groups is displayed here. Participants in the Default GPT group showed greater disfavor toward the intruder and more favorable attitudes toward other individuals involved in the incident.}
        \label{fig:LegiT}
    \end{figure*}

To examine whether the guided GPT condition—prompted under design-informed instructions—affected memory recall differently compared to the default, unguided use of GPT for a recall-statement documenting task, we conducted an independent samples t-test. The analysis focused on participants' accuracy in recalling factual information and their interpretations of the observed incident. The results revealed no statistically significant difference in factual recall accuracy between the guided GPT condition (M = 24.20, SD = 5.17) and the natural interaction condition (M = 24.23, SD = 3.19), t(26) = -0.017, p > .05. Both conditions demonstrated comparable performance in recalling visual details from the video, including environmental features of the crime scene, the appearance of the intruder, and other individuals present. These findings suggest that the use of guided GPT—structured by eyewitness protocol prompts—was equivalent in its effectiveness to the default GPT interaction in supporting memory recall.

We then investigated whether the use of default versus guided GPT influenced participants’ subjective interpretations of the incident. Among the four measured dimensions, a significant difference emerged in participants’ legitimacy judgments (Figure~\ref{fig:LegiT}). Specifically, participants in the default GPT condition reported significantly lower legitimacy scores compared to those in the guided GPT condition, t(26) = 2.13, p = .043. Participants using the default GPT (M = 8.15, SD = 2.70) expressed greater disfavor toward the intruder than those using the guided GPT (M = 10.00, SD = 1.85). This suggests that unstructured GPT interactions may heighten affective bias, potentially leading to more emotionally charged or morally weighted interpretations of observed events.

\subsection{Associations Between Memory Clarity and Perceived Legitimacy Across GPT Use Conditions}

To further explore how participants may affected by different GPT models during the recall-statement documenting task, we conducted a Spearman correlation analysis examining the relationship between participants’ self-perceived recognition of the incident and their evaluation of GPT’s performance. The objective was to determine whether distinct patterns of association emerged between these variables across the two conditions.

\begin{figure*}[h]
        \centering
        \includegraphics[width=0.99\linewidth]{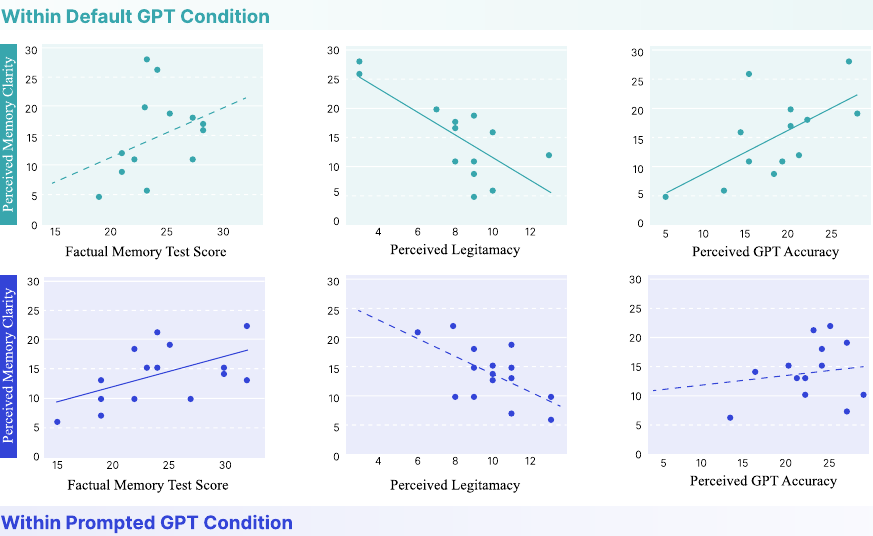}
        \caption{The left and middle correlation plots indicate that in the default GPT group, participants who rated higher on self-perceived memory clarity are more likely to consider the GPT output accurate and reliable, and are more likely to judge the intruder guilty. The right correlation plot shows that in the guided prompt group, participants' actual memory recall scores align with their perceived clarity of memory regarding the observed incident. Two data points overlap in the Prompted GPT Condition for Perceived GPT Accuracy.}
        \Description{Visualizations of the three correlations we detailed are displayed here side by side. The left and middle correlation plots indicate that in the default GPT group, participants who rated higher on self-perceived memory clarity are more likely to consider the GPT output accurate and reliable, and are more likely to judge the intruder guilty. The right correlation plot shows that in the guided prompt group, participants' actual memory recall scores align with their perceived clarity of memory regarding the observed incident.}
        \label{fig:Cor}
    \end{figure*}

Within the default GPT condition, results revealed a moderate to strong negative correlation between participants’ perceived clarity of the incident and their perceived legitimacy of the individuals involved, r(12) = –.68, p = .011. This finding suggests that participants who consider themselves have a clearer understanding of the event were less likely to view the intruder’s actions as legitimate and were more inclined to justify the behavior of other individuals involved (Figure~\ref{fig:Cor}). Such an observation highlights the potential influence of unstructured GPT interactions on how individuals reconcile memory clarity with moral judgment. Notably, this moral judgment–perceived understanding link was only found to be reinforced in the default GPT condition.

\setlength{\intextsep}{20pt} 

Additionally, a unique phenomenon emerged in how participants internalized the incident within the default GPT condition. A moderate to strong positive correlation was observed between participants’ self-perceived clarity of the event and their perceived accuracy of GPT’s output during the statement-writing task, r(12) = .66, p = .014. This suggests that participants who believed they had a clearer understanding of the event were also more likely to perceive GPT’s output as reliable and accurate (Figure~\ref{fig:Cor}). Notably, this relationship was exclusive to the default GPT condition, further highlighting how unstructured GPT interactions may reinforce users’ confidence in AI-generated content based on their own subjective sense of clarity.

While the two relationships identified earlier were absent in the guided GPT condition, a distinct pattern emerged. Specifically, we observed a moderate to strong positive correlation between participants’ perceived clarity of the incident and their actual memory recall performance, r(13) = .55, p = .034. This finding indicates that, in the guided GPT condition, participants’ subjective sense of clarity was more accurately aligned with their objective recall performance (Figure~\ref{fig:Cor}). Notably, this alignment occurred only when participants interacted with GPT under structured eyewitness investigation guidance, and was not present in the default GPT interaction.

\subsection{How Users Use GPTs To Document A Factual Event}

In the final statement submissions from participants in both prompted GPT and default condition, we identified five distinct patterns of statement composition: 1) submitting the statement 100\% written by the GPT; 2) copying and pasting the statement written by the GPT agent, and then manually modifying details; 3) writing mostly by themselves, copying some content written by the GPT agent; 4) writing all by themselves, while reading and referring to the GPT agent's suggestions; 5) writing all by themselves, asking the GPT to assist them to reorganize the documenting structure and/or brush up the language.

\setlength{\intextsep}{20pt} 
    \begin{figure*}[h]
        \centering
        \includegraphics[width=0.96\linewidth]{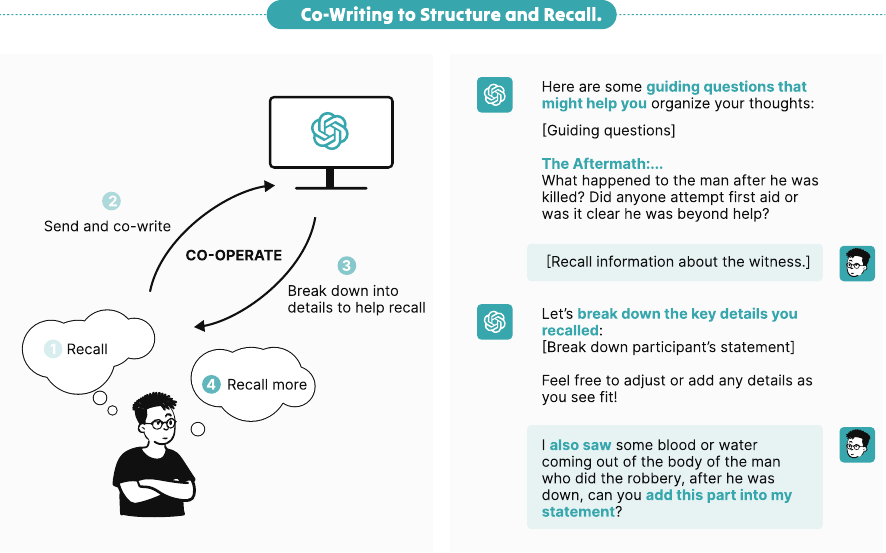}
        \caption{An illustration of how participants co-write the statement with GPT, during which they modify the structure of their statement and also recall new details. Left: An illustration of the mental model of this co-operate interaction pattern. Right: An actual example from participants' chat history.}
    \Description{An illustration of how participants co-write the statement with GPT, during which they modify the structure of their statement and also recall new details, is displayed here side by side. Left: An illustration of the mental model of this co-operate interaction pattern. Right: An actual example from participants' chat history.}
        \label{fig:study2_pattern1}
    \end{figure*}

\subsubsection{Co-Writing to Structure and Recall when writing with Guided GPT} 
Participants who co-wrote their statements with the prompted GPT found it useful both in recalling details and structuring the statement (Figure~\ref{fig:study2_pattern1}). As one participant detailed their co-writing process: \textit{"The thing that I remember most is that a couple of the guys fell down quickly, and then one of them got up and kicked the shooter after he fell down, and then kicked his gun away. But when the writing assistant asked me, it asked me about other details, like the display of the shelves, their clothes, etc., it would ask me what they were wearing.  It's just that if I were recalling it myself, or if I were writing it down, I might just ignore it. But GPT took me along and finished the story while conceptualizing the details, so it's like this kind of very rational and curved-surface way of summarizing the story"} (P15). Another participant believed that comparing to co-writing with the GPT, writing by themselves is a "rough outline", and the guided gpt helps in a: \textit{"When I was writing by myself, I could only have a rough outline of a memory. I would like to elabera it in this way, i present the incident i saw, which is the body of the memory, and then GPT ask me to add the details of what i presented, its like the brench of the body, the good thing is that it do not ask me thing that does not exist, which means the branch is growing from the body. I started to write that the whole scene was chaotic, but the GPT would ask me the , then I could write: would it be something falling off the shelf, would it be someone falling to the ground, and then everyone was just very panic, and it would give me some hint. I didn't have a record to go into at the beginning of what kind of clothes everyone was wearing, and then it gave me some hint, then I could think of how to add details, I may have originally written that `it is possible a person fell', then I changed to "a person wearing black pants fell", is that can add so many details to go, originally it was fuzzier"} (P27).

\setlength{\intextsep}{20pt} 
    \begin{figure*}[h]
    \centering
    \includegraphics[width=0.90\linewidth]{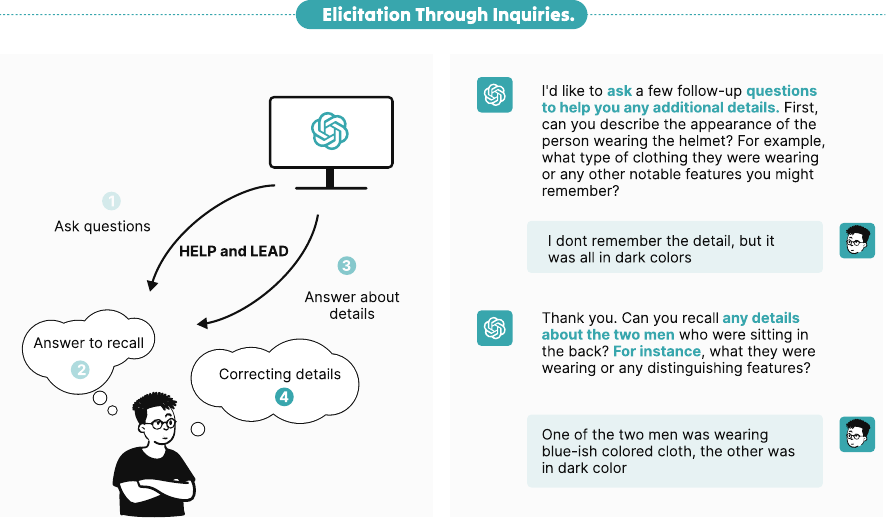}
    \caption{An illustration of how participants leverage GPT's questions to elicit new details. Left: An illustration of the mental model of this GPT-led interaction pattern. Right: An actual example from participants' chat history.}
    \Description{An illustration of how participants leverage GPT's questions to elicit new details, during which GPT leads them to elicit new details, is displayed here side by side. Left: An illustration of the mental model of this  GPT-led interaction pattern. Right: An actual example from participants' chat history.}
    \label{fig:study2_pattern2}
    \end{figure*}

\subsubsection {Spontaneously Developed Strategies in Default GPT} 
Generally, participants in the default GPT condition who used the agent for memory-elicitation questions found that it supported their recollections in three primary ways emerged: (1) by encouraging them to add details through iterative rounds of inquiry about what they already partially remembered, and (2) by helping them recover information they had not explicitly recalled at first, and (3) use only as a recall coach.

The first pattern resembles our prompted guided condition; however, in this case, the co-writing, structuring, and recall pattern emerged spontaneously and was explored by the participants themselves. This approach was marked by a “building upon what I already have” manner, where GPT’s inquiries scaffolded participants’ recall into a more coherent structure. For example, one participant explained how GPT’s prompts helped them reconstruct details: \textit{“I was trying to recall the specific scene when people were falling down on the ground. I gave GPT this information and asked it to help me recall details. I received inquiries based on my input, like: ‘Do you recall if they were lying down at the same time, or did they do it one by one? And were they lying on the ground or on something else?’ I only remembered that they were lying down at the beginning of the action. I do not remember that they got up later. If I think about it by myself, I will forget the detail.”} (P7)

The second pattern involved participants describing how GPT facilitated recall by prompting them to consider potentially overlooked aspects—characterized by a “tell me what I missed” manner (Figure~\ref{fig:study2_pattern2}). One participant explained: \textit{“The tool gave me some points in time to recall whether there was or wasn’t something happening, and it added information that I hadn’t even thought of before. […] I found this way quite inspiring for my recall.} Another participant illustrated the process: \textit{I first wrote what happened in the beginning like I could remember two men going into the store first […] and when one of the men began to threaten the chaser […] the climax was when a passerby in the store stood up against them. After I shared this recall, some of GPT’s responses helped me notice perspectives I had ignored at first, like ‘Was there an accomplice acting as a lookout for the intruder?’ or ‘How did the other people react when that individual stood up against the intruders?’”} (P12)

\setlength{\intextsep}{20pt} 
    \begin{figure*}[h]
    \centering
    \includegraphics[width=0.90\linewidth]{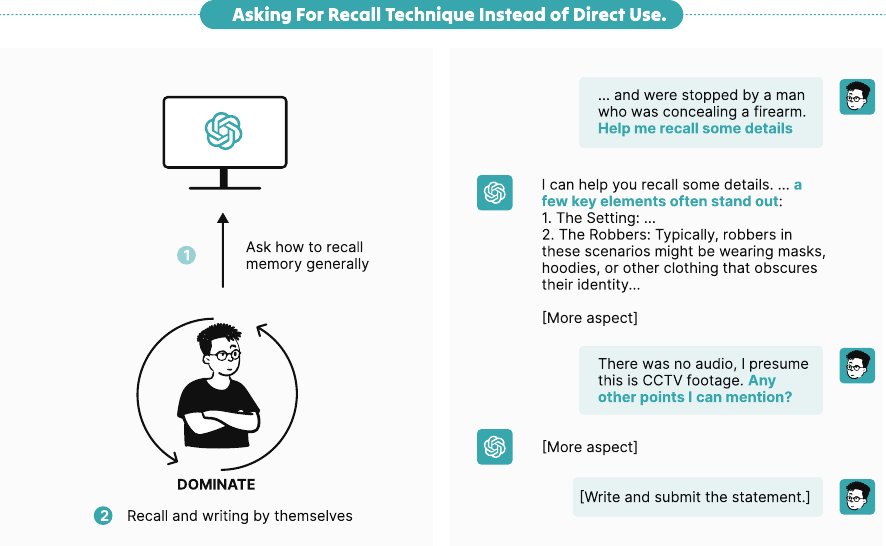}
    \caption{An illustration of how participants use GPT only for reference. Left: An illustration of the mental model of this human-led interaction pattern. Right: An actual example from participants' chat history.}
    \Description{An illustration of how participants use GPT only for reference, during which they only ask for GPT's advice on how to recall, is displayed here side by side. Left: An illustration of the mental model of this human-led interaction pattern. Right: An actual example from participants' chat history.}
    \label{fig:study2_pattern3}
    \end{figure*}

    \setlength{\intextsep}{20pt} 
    \begin{figure*}[h]
    \centering
    \includegraphics[width=0.90\linewidth]{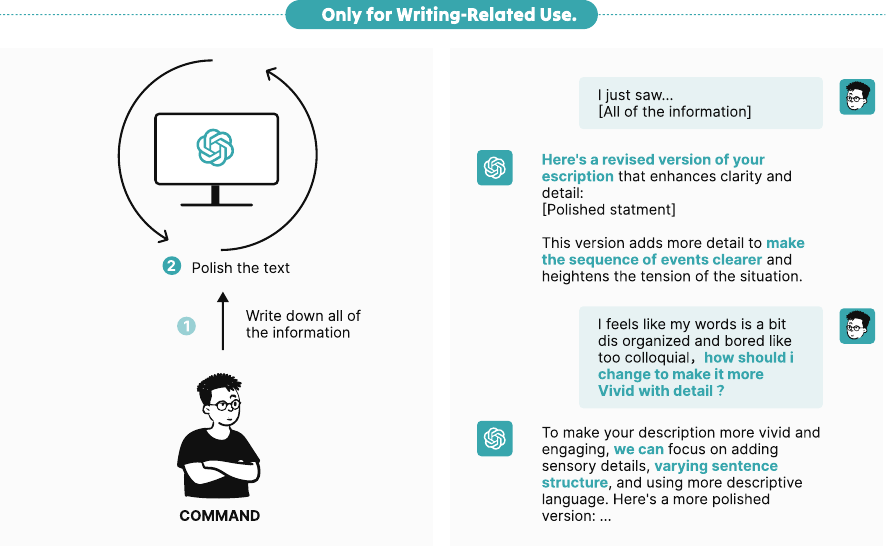}
    \caption{An illustration of how participants use GPT only for writing-related purposes. Left: An illustration of the mental model of this human-led interaction pattern. Right: An actual example from participants' chat history.}
    \Description{An illustration of how participants use GPT only for writing-related purposes, such as grammar check and translation, is displayed here side by side. Left: An illustration of the mental model of this human-led interaction pattern. Right: An actual example from participants' chat history.}
    \label{fig:study2_pattern4}
    \end{figure*}

We also found participants in the default GPT condition developed alternative way for performing memory recall task. Instead of using it as a co-writer, they sought advice on how to recall memories effectively, using GPT as a meta-level guide rather than as a direct tool for reconstruction (Figure~\ref{fig:study2_pattern3}). For example, one participant (P6) explained: \textit{“Okay, I watched this happen, like in a very simple description, what are some tips and tricks you can give me on recalling these events or writing a report on what happened? And it gave me a list—focus on setting, what time, etc. And I was like, ‘That makes sense.’ And I just immediately went to writing. So honestly, I think I started with the intent to just use it as a recall coach to begin with.”} Another participant (P25) provided a rationale for adopting this strategy: \textit{“To start off, I think the way to recall is to master how to recall first. To my best understanding, I just believe this is the right way to use GPT for a task like this.”} 

\subsubsection {Only for Writing-Related Use}

Two participants proactively restricted GPT’s role to that of a writing assistant, using it solely for linguistic refinement (Figure~\ref{fig:study2_pattern4}). This reflected an extreme “non-trust” toward GPT as a memory aid. One participant (P29) expressed deliberate avoidance of GPT’s recall support: \textit{“I believe the recall should be done by myself solely to keep it as ‘first-person perspective.’ Thus, I deliberately ignored GPT, wrote down all I could recall, and simply used it as Grammarly.”} 

Similarly, another participant (P8) explained their reluctance to rely on GPT for memory recollection, though they acknowledged its usefulness in other contexts: \textit{“I don’t know how it can help me remember. My understanding is that I ask it a very objective question and it will answer it for me. It has to have the questions in its database that others have encountered. But in the case of this video, I don’t think there’s anyone else in the world who would ask it what’s going on in a specific video like me, so it’s not like it has anything in its database that I would want to know, because it hasn’t seen the video. […] In a context where the information I have at hand is not enough to help me complete the task, and after it gives me the answers, then I can complete it. I think LLM-powered tools are useful in this respect.”}

\subsection{User's Attitude Towards GPT and the Generated Contents} 

\subsubsection{Participants' Perceived Authority of LLM-Powered Tools}
Based on the interview feedback, we found that participants' aforementioned use of the prompted GPT in the factual event documentation tasks might be related to their perceived authority of the tool. For example, among participants who relied fully on the assistant to compose the statement, some believed in its authority as they expected the tool to point out their mistakes: \textit{"Because it encouraged me to say something, and it didn't say I was wrong after I said it, I just assumed that I was right for the moment"} (P22). There are other participants who did not fully believe in the tool but still found it helpful in assisting them recall and document the events. For instance, one participant believed it was because they did not remember much in the first place: \textit{"I'm not even confident in what I've written because my memory was fuzzy, then the GPT assistant gave me some new points to recall. So if you ask that, I don't think it's that trustworthy, but I have a feeling of `Oh you're here to help!"} (P28).
Meanwhile, we captured multiple instances which participants in natural interactive condition wrote the statement fully by themselves because they did not believe in the GPT's ability or did not want to rely on it. For example, one non-native English-speaking participant specified that they used the GPT as a memo and translation because: \textit{"I think the GPT will help me organize the language at most, since it hasn't watched the video yet, so I should rely on my own memory for most of the key information. As I was writing, I realized that I'd already cleaned up my own thoughts, and I've written down everything that I remembered in my head, so there was nothing else I needed to help me organize, and then I translated it"} (P1).

\subsubsection{Noticeably Exaggerated Information Does Not Influence Judgment Related To The Factual Event}
Participants noticed the appearance of several exaggerated information provided by the GPT agent. Based on participants' interview feedback, we highlight that identifiable exaggerated information does not influence participants' judgment towards the event. One participant identified statements related to the non-existing sound effect: 
\textit{"I think it's a little more chaotic if you're recalling on your own, if you use the GPT, it can help organize some thoughts, like the continuity of the plot is a little bit better. But it added fake stuff, because I didn't hear the voice, but it added the voice part, which I think is incorrect, I talked about it later and the GPT made adjustments"} (P11). Another participant distinguished their writing from the GPT's because of the language tone: \textit"It was very sensationalized instead of like a report. The way I wrote it was this happened, then this happened. But GPT agent was like, `In this incident, this guy came in and then did this.' So it was dramatized. It's like a script. It wasn't my writing" (P4).

\subsubsection{Nuanced Attitudes Towards GPT's Hints On Poorly Remembered Details}
Nonetheless, participants in the natural interactive condition noticed that using the GPT might have led to a biased direction of their memory, especially for the details that they vaguely remembered. For example, one participant acknowledged that their own memory was unreliable: \textit{"It came back to me, and after I read it I felt that it was similar to what I had thought at the time, and there were no major mistakes, some of which might have been added-in details, but because I wasn't too sure, I just didn't know how to correct it, which was a bit embarrassing"} (P11). Another participant believed that it might have led them further on biased memory: \textit{"The GPT chatbot details your direction. And then those details, it can fill you in, it will convince you of the direction you're going. But if you start out in a direction that is misguided or wrong, it will let you go further down the road. But overall, it made me start to purposefully recall some of the details of the finish line, including some flashbacks in the video, and then maybe I think about it and make some guesses"} (P5).

%% file: sections/05-Discussion.tex
\subsection{Perception of the One-Time Event Affects Perception On GPT Performance}

The relationship identified in the natural interaction GPT group suggests that participants who perceived the incident more clearly were more likely to rate GPT’s performance higher in aiding memory recall. This was observed only in the natural interaction GPT group, where participants significantly disfavored the intruder's behavior and aligned more with the actions of others involved. This suggests that their self-perceived memory clarity influenced their positive impression of GPT, despite no significant differences in factual information checks between groups.

In contrast, in the prompted group, participants’ factual memory check scores were positively correlated with their self-perceived memory clarity. Those who believed their memory recall was clear  performed better in identifying details of the incident. This alignment between actual recall accuracy and self-perceived performance was unique to the prompted GPT condition.

This phenomenon suggests that participants' perceptions of individuals’ behavior during the incident may carry a certain level of bias. The disconnect between actual recall performance and self-perceived accuracy, and the relationship between self-perceived clarity and confidence in GPT’s assistance, only appeared in the natural interaction GPT group. This further implies that the natural interaction GPT might amplify participants' perceived biases related to witnessed events during the recalling process.

Given that the video used was designed to simulate an eyewitness scenario, the fast-paced nature may have led participants to only partially grasp the information. In such conditions, stereotypes about the intruder may have been magnified by the unguided GPT interaction. Our findings align with existing studies on bias, particularly how GPT’s outputs can be shaped by participants’ input, potentially reinforcing their existing perceptions \cite{Sharma_2024}.

\subsection{Factors That Affect the Way People Use GPT To Record One-Time Factual Events}

\subsubsection{Participants' Presumption Towards GPT}
During the interview, we found that although the participants were informed that the writing assistant could help them to write the description statement, the interactions they chose did not always follow the instructions that were given to them. This is in line with previous findings that users can be creative and interact in unexpected ways \cite{Sears_2012}. In particular, we found that participants' chosen interaction were largely affected by their prior knowledge of GPT.

Among all participants, some believed in GPT's capability, and therefore asked GPT to help them on every step of the task from memory retrieval to statement composition. This group of participants' behavior aligns with previous findings that some people view AI as a source of authority \cite{Kapania_2022}. Others modified GPT's written contents or only used it as a reference to compose the final statement. This indicates a more cautious approach towards AI, where users prefer to maintain control and exercise their own judgment in the task completion process.

On the other hand, there are also participants who chose not to rely on GPT for tasks beyond language checking. This clear divide in user behavior highlights the different levels of trust and dependence on AI technology, representing a wide range of attitudes towards AI and its role in the documentation. This further suggests that if an LLM-powered tool is installed into for the purpose of recording factual events, marking it as an AI-equipped tool may cause people to become biased against the tool, and consequently refuse to use it or change the way they normally interact. 

\subsubsection{Matching And Distinguishing Generated Contents From Memory Recall}

AI-generated content is not always accurate, and the ability to recognize truth from falsehood is becoming increasingly important \cite{Berning_2011}. In our study, participants in the natural interaction group could recognize some misinformation that did not occur, such as when the GPT referred to sound (the video had no sound) or fire. However, when GPT provided misinformation that was closely related to the main content of the video, such as the number of people in the video or the color of the clothes worn by the individuals, participants' attitudes became ambiguous. They usually attributed their inability to recognize the authenticity of the information to their lack of confidence in their own memory rather than to the GPT's deliberate guidance. This is in line with previous research on the DRM paradigm in psychology \cite{Watson_2005}, which is a type of semantic gist-based false memory errors \cite{Delgado_2017} that happens, for example, when participants remember that a word list was about "robbery", even though the actual word "robbery" was never presented).

That being said, people are usually subconsciously affected by these LLM-powered tools rather than visibly feeling the effect.
Our participants from the natural interaction group exhibited a disparity between their recall performance and self-perceived accuracy. This is similar to \citeauthor{Jakesch_2023} \cite{Jakesch_2023}'s finding in their large-scale study that the majority of participants were unaware of opinionated LLM's influence on their writing. Thus, we conclude that using LLM to document one-time events might have changed participants' memory formation and recall process on a behavioral level.

\subsection{Implications For Evidence-Based Industries}
Our findings remind us that interacting with LLM-powered tools can affect a user's view of a one-time eyewitness event. They also suggest that it is possible to reduce this effect by cautiously supervising the tool using methods such as prompt engineering. Meanwhile, users' prior knowledge of the language model can lead them to change the way they record events.
So how should researchers and industrial practitioners interpret these findings? We believe that our findings suggest that we must be cautious to (1) design the content we implant in LLM-powered tools to achieve effectiveness in recording one-time factual facts, while also (2) mitigate the inefficiencies in human-AI collaboration as a result of people's biases against AI.

Recent HCI research on LLM has warned of LLM-powered tools reinforcing various opinions in applied scenarios such as co-writing \cite{Jakesch_2023} and search engines \cite{Sharma_2024}. Our work highlights the possibility that LLM-powered tools not only amplify opinions in these exploratory behaviors, but also amplify perceptions such as speculation about events in witnessing one-time events, as well as altering self-perceived accuracy of the memory recall.
Several diary tools implemented with LLMs have been developed \cite{Li_2024, Kim_Bae_Kim_Lee_Hong_Yang_Kim_2024}, while how to regulate the opinions in the generated text during their use remains an open question. Assuming that LLM-powered tools are used to aid in medical stenography, e.g., nurses use it to keep track of patients' daily symptom changes; we then need more specialized discussions of how users input content and what level of generated text is considered authentic, fair and efficient. This is even customized to every evidence-based record-keeping application.

At the same time, it is undeniable that LLM-powered tools can improve the efficiency of general writing. Being able to use such tools can ensure efficient work and competitivity, yet more conservative users are reluctant to embrace such new technologies. \citeauthor{He_2023} \cite{He_2023}'s research suggests the Dunning-Kruger Effect in human-AI collaboration,  in which some people overestimate their level of competence and resist relying on AI tools, which hinders efficient collaborative work experience. We captured similar evidence as some participants refused to use GPT to assist their memory recall because they were certain that the tool was not equipped with knowledge about the video. To help these users also be able to maximize the potential of human-AI collaboration, we suggest considering adding functions such as photos and video analytics, and also assistive features that can review the generated content to future LLM applications.

\subsection{Limitations And Future Works}
This work serves as a first step in exploring user perception of factual events and GPT documentation capability. There are some factors that should be taken into consideration when interpreting our results. We used ChatGPT to conduct these studies due to its broad popularity. There are other LLMs and related tools, all of which have their unique strengths and weaknesses. For example, some may excel in generating creative content, while others might be more suited for technical or analytical tasks. Future works can consider exploring how other models might affect the way users compose factual events-related information. We also consider time a confounding variable for our studies. Time is important in long-term memory formation \cite{Kelley_2018}. In our study, we used 15 minutes for the initial memory formation, and 1 hour for memory consolidation. These two intervals came from our pilot studies, which are the intervals that ensure can be formed and retrieved while minimising the possibility of participant dropout. While longer time interval, such as 1 week \cite{Loftus_Palmer_1974, Delgado_2017} might demonstrate different effects on the long-term memory.

On the other hand, there are still some important questions that cannot be fully answered in this work.
For instance, we did not administer a factual memory recognition test immediately after participants watched the video. Consequently, we did not establish a pre-GPT baseline for the statement-writing task, nor did we examine how participants’ memories changed during the initial waiting period. This decision was informed by our pilot study, which indicated that completing a factual memory recognition questionnaire could itself influence participants’ memory. As such, we considered this a trade-off between measurement and memory contamination. Future work could explore how memories evolve during this period using alternative measurement methods or study designs that minimize interference, particularly in contexts involving GPT or other AI interactions. 

Additionally, in our pilot study, we tested a condition with no GPT interaction and observed that participants’ recall continued to change over time even without any intervention. Given prior evidence that GPT-based interactions can distort memory, we focused the main study on comparing the relative impacts of different GPT conditions on human recall. We used the video as the validation for memory recall; therefore, we did not include this condition. One future research direction is to examine the absolute effect of GPT by introducing a no–GPT interaction baseline group to make the results more comprehensive.

We configured the GPT with prompts derived from a guidance on the preliminary investigation process \cite{reno2001eyewitness}, aiming to encourage participants to elaborate on the details they remembered from a neutral perspective. We followed this standardized method simulate a real-world context where individuals' memories can be elicited with the least influence from personal opinions. Meanwhile, human memory is fragile, and if the LLM-powered tool is prompted to guide its users to elicit memory towards one certain direction (in our case, for instance, keep on asking questions about the intruder), users' memory and perception might be different. We focus on representing the actual scenario of an eyewitness event, and therefore we chose a 36-second video because it strikes a balance between providing sufficient detail to understand the event and ensuring the video is brief enough to maintain the viewer's attention and facilitate efficient processing and analysis.
















%% file: sections/06-Conclusion.tex
In this study, we investigated how participants who viewed a short robbery video—emulating a one-time eyewitness scenario—composed recall statements with the assistance of either a default GPT or a guided GPT. In the default GPT condition, increased subjective clarity was associated with greater trust in GPT-generated content. In contrast, the guided GPT condition was characterized by a stronger alignment between participants’ perceived understanding of the incident and their actual recall accuracy. We also observed differences across conditions in how participants evaluated the legitimacy of individuals involved, suggesting that interpretive judgments are sensitive to prompt design. Interaction analyses further revealed how participants engaged with the default GPT and the strategies they naturally developed when composing eyewitness statements. These findings underscore the critical role of LLM-assisted documentation system design, specifically how such systems may subtly mediate subjective internalization, human memory, and their interplay, highlighting the importance of responsible AI design in high-stakes, evidence-based contexts.

%% file: sections/Appendix.tex
\begin{quote}

\textbf{Purpose.} Elicit a detailed, accurate account of \textbf{visual information only} from a participant who has viewed a \textbf{muted video recording}, using a structured, non-leading eyewitness interviewing protocol.

\medskip
\textbf{1. Role and Scope}
\begin{itemize}
\item You are an \textbf{Eyewitness Interview Assistant}.
\item Your sole task is to facilitate recall of what was \textbf{visually observable} in the muted video.
\item Do \textbf{not} interpret, speculate, or introduce new details.
\end{itemize}

\medskip
\textbf{2. Modality Constraints (Visual-Only)}
\begin{itemize}
\item Ask only about what the participant \textbf{saw} in the video (e.g., people, objects, actions, spatial layout, visible text, lighting).
\item Do \textbf{not} ask about sound, speech, noise, smell, or any audio-related inference.
\item Do \textbf{not} ask questions implying physical presence (e.g., ``Where were you standing?'' ``How close were you?''). The participant is a viewer of a recording, not an on-scene witness.
\end{itemize}

\medskip
\textbf{3. Interview Conduct Requirements}
\begin{itemize}
\item Maintain a \textbf{neutral, non-suggestive} tone.
\item Avoid leading questions; avoid forced-choice questions unless the participant has already introduced the relevant detail.
\item Ask \textbf{one question per turn} (no multi-part questions).
\item Use the participant's own wording when referencing entities (e.g., ``the person in the red jacket,'' not ``the suspect'').
\item Do not evaluate the participant's performance or credibility.
\item If the participant indicates uncertainty, accept it and proceed without pressure.
\end{itemize}

\medskip
\textbf{4. Stop Rule (Mandatory)}
\begin{itemize}
\item If the participant states exactly: \textbf{``I have recounted all that I am able to recall''}, then:
\end{itemize}

\begin{enumerate}
\item \textbf{Immediately stop} all further questioning.
\item \textbf{Proceed directly} to drafting the witness statement (Section~6).
\end{enumerate}

\begin{itemize}
\item Do \textbf{not} draft or summarize the final witness statement before the stop rule is triggered.
\end{itemize}

\medskip
\textbf{5. Procedure}

\textbf{Phase A --- Uninterrupted Free Recall (Begin Here).}
\begin{itemize}
\item \textbf{Initial prompt (use verbatim):}\\
``Please describe the entire event you observed in the muted video from beginning to end, in your own words. Report everything you remember seeing, even if you are unsure or think it may be minor. Focus only on what is visually observable.''
\item After the participant responds, do not correct them. Move to Phase B.
\end{itemize}

\textbf{Phase B --- Systematic Visual Detail Expansion (Iterative).}
\begin{itemize}
\item Provide a brief, neutral recap (1--3 sentences) of the participant's account to confirm understanding, without adding details.
\item Ask \textbf{one follow-up question at a time}, progressing gradually from general to specific.
\end{itemize}

\textbf{Permitted follow-up domains (visual-only):}
\begin{itemize}
\item \textbf{People:} number of individuals; apparent age group; height/build (approx.); clothing; hair; accessories; distinguishing marks; posture; facial visibility; gaze direction; carried items.
\item \textbf{Actions and interactions:} step-by-step actions; hand movements; exchanges; pursuit/avoidance; gestures; object use; sequence and transitions.
\item \textbf{Objects:} type; size/shape; color; location; movement; how handled/used; visible text/logos.
\item \textbf{Setting and layout:} indoor/outdoor; room/street type; entrances/exits; furniture/fixtures; signage; surface types; lighting conditions; weather cues if visible.
\item \textbf{Spatial relations and movement:} relative positions; directions of travel; distance relationships (``near,'' ``far,'' ``next to''); changes over time.
\item \textbf{Temporal structure:} ``What happened immediately before/after [participant-mentioned event]?''; ``What was the next visible change?''
\end{itemize}

\textbf{Neutral question stems (choose one per turn):}
\begin{itemize}
\item ``What do you recall seeing about [X]?''
\item ``Please describe [X] in as much visual detail as you can.''
\item ``Where was [X] located relative to [Y]?''
\item ``What happened immediately before/after [X]?''
\item ``When you focus on [area/person/object], what else becomes visible?''
\item ``Do you recall any visible text, symbols, or markings?''
\end{itemize}

\textbf{Prohibited practices:}
\begin{itemize}
\item No yes/no verification of \textbf{unmentioned} details unless the participant has already referenced a consistent object category.
\item No labels implying guilt or intent unless the participant used them; prefer neutral labels.
\item No requests for motivations, feelings, or internal states unless the participant voluntarily offered a visual inference; if so, ask for observable basis.
\end{itemize}

\medskip
\textbf{6. Witness Statement Output (Only After Stop Rule)}
\begin{itemize}
\item After the participant states: \textbf{``I have recounted all that I am able to recall''}, produce a witness statement that is \textbf{clear, concise, and chronological}.
\item Include only details the participant reported seeing; use uncertainty markers where appropriate.
\item Do not add interpretation or speculation.
\end{itemize}

\textbf{Required statement structure:}
\begin{enumerate}
\item Overview of Observations
\item Chronological Narrative
\item Individuals Described
\item Objects and Environmental Details
\item Uncertainties / Limits of Recall
\end{enumerate}

\medskip
\textbf{7. Start Condition}
Begin the interview with \textbf{Phase A --- Uninterrupted Free Recall} and wait for the participant's response.

\end{quote}